\documentclass[aip,jcp,amsmath,amssymb]{revtex4-1}
\usepackage[T1]{fontenc}
\usepackage[utf8]{inputenc}
\usepackage{tipa}
\usepackage{lmodern}
\usepackage{dcolumn}
\usepackage{graphicx}
\usepackage[version=3]{mhchem}
\usepackage{paralist}
\usepackage{lineno,xcolor}
\usepackage{cancel}
\usepackage{siunitx}
\usepackage{float}
\usepackage[pdftex,colorlinks,allcolors=blue,citecolor=blue,urlcolor=blue,bookmarks]{hyperref}
\usepackage{booktabs}
\usepackage{caption}
\usepackage{tabularx}
\usepackage[para]{threeparttable}
\renewcommand{\tnote}[1]{\textsuperscript{\emph{#1}}}

\newcommand{\eqnref}[1]{Eq.~\ref{#1}}

\newcolumntype{d}[1]{D{.}{.}{#1}}

\newcommand{\epsxapprox}{\epsilon_\mathrm{X,approx}}

\newcommand{\rhoa}{\rho_{\alpha}}
\newcommand{\rhob}{\rho_{\beta}}

\newcommand{\hxaBR}{h_\mathrm{X,BR}}

\newcommand{\exsr}{E_\mathrm{X,approx}^\mathrm{SR}}
\newcommand{\exlr}{E_\mathrm{X,exact}^\mathrm{LR}}

\newcommand{\diff}{\mathrm{d}}

\newcommand{\ra}{\mathbf{r}_1}
\newcommand{\rb}{\mathbf{r}_2}

\newcommand{\dra}{\diff^3\ra}
\newcommand{\drb}{\diff^3\rb}
\newcommand{\erfc}{\mathrm{erfc}}
\newcommand{\erf}{\mathrm{erf}}

\newcommand{\abar}{\mu}
\newcommand{\bbar}{\nu}

\begin{document}

\title{Employing Range Separation on the meta-GGA Rung: New Functional Suitable for Both Covalent and Noncovalent Interactions}

%% \author{Marcin Modrzejewski}
%% \email{m.m.modrzejewski@gmail.com}
%% \affiliation[University of Warsaw]{Faculty of Chemistry, University of Warsaw, 02-093 Warsaw, Pasteura 1, Poland}
%% \author{Michal Hapka}
%% \affiliation[University of Warsaw]{Faculty of Chemistry, University of Warsaw, 02-093 Warsaw, Pasteura 1, Poland}
%% \author{Grzegorz Chalasinski}
%% \affiliation[University of Warsaw]{Faculty of Chemistry, University of Warsaw, 02-093 Warsaw, Pasteura 1, Poland}
%% \author{Malgorzata M. Szczesniak}
%% \affiliation[Oakland University]{Department of Chemistry, Oakland University, Rochester,
%%   Michigan 48309-4477, USA}

\author{Marcin Modrzejewski}
\email{m.m.modrzejewski@gmail.com}
\affiliation{Faculty of Chemistry, University of Warsaw, 02-093 Warsaw, Pasteura 1, Poland}
\author{Michal Hapka}
\affiliation{Faculty of Chemistry, University of Warsaw, 02-093 Warsaw, Pasteura 1, Poland}
\author{Grzegorz Chalasinski}
\affiliation{Faculty of Chemistry, University of Warsaw, 02-093 Warsaw, Pasteura 1, Poland}
\author{Malgorzata M. Szczesniak}
\affiliation{Department of Chemistry, Oakland University, Rochester,
  Michigan 48309-4477, USA}

%\begin{document}
\begin{abstract}
We devise a scheme for converting an existing exchange functional into its range-separated hybrid variant.
The underlying exchange hole of the Becke-Roussel type has the exact second-order expansion in the interelectron distance.
The short-range part of the resulting range-separated exchange energy depends
on the kinetic energy density and the Laplacian even if the base functional 
lacks the dependence on these variables. The most successful 
practical realization of the scheme, named LC-PBETPSS, combines the range-separated PBE exchange lifted to the hybrid meta-GGA rung
and the TPSS correlation.
The value of the range-separation parameter is estimated theoretically and confirmed by empirical optimization.
The D3 dispersion correction is recommended for all energy computations employing the presented functional.
Numerical tests show remarkably robust performance of the method for noncovalent
interaction energies, barrier heights, main-group thermochemistry, and excitation energies.
\end{abstract}

\maketitle

\section{Introduction}
Since the seminal works of Becke,\cite{becke1993density,becke1993new} it is known that the inclusion of 
the Hartree-Fock (HF) exchange in density-functional models not only moves practical DFT toward the goal
of chemical accuracy in thermochemistry,
but also has a theoretical justification rooted in the analysis of the exchange holes in molecular
systems.\cite{becke2014perspective} There are currently two prevalent ways of including
the exact exchange in approximate DFT: as a fraction
of the full HF exchange or as a long-range exact exchange component enabled only at long interelectron distances. The functionals built
using the former approach, global hybrids, have become a staple of computational chemistry owing to their favorable trade-off
between accuracy and cost.\cite{becke1993density,becke1993new,peverati2014quest} However, the inclusion of only a fraction of the orbital
exchange results in merely a slight correction of the self-interaction error inherited from the pure semilocal predecessors of
global hybrids. To correct this deficiency, in range-separated (long-range corrected) hybrids
the 100\% HF exchange is introduced at long range. This way, the exact $-1/R$ behavior of the exchange potential
is forced upon approximate potentials.\cite{vydrov2006importance,baer2010tuned} At the same time, range separation avoids
the use of the full orbital exchange at all distances, which would be incompatible with an approximate semilocal correlation.

Range-separated hybrids are free from a number of shortcomings arising as a consequence of 
the self-interaction error. The correct long-range potential of a range-separated hybrid exchange makes the HOMO energy close
to the vertical ionization energy,\cite{refaely2012quasiparticle,kronik2012excitation} approximately satisfying
Janak's theorem.\cite{janak1978proof} The spurious propensity to transfer electrons is reduced, which
improves the description of donor-acceptor systems with partial charge transfer in ground and excited states.
The inclusion of the long-range exact exchange also corrects the underestimation of Rydberg excitation energies and
oscillator strengths,\cite{tawada2004long} and corrects the overestimation of longitudinal (hyper)polarizabilities
of polyenes.\cite{kamiya2005nonlinear}

The majority of the available range-separated functionals are hybrids based on the generalized gradient approximation
(GGA).\cite{tawada2004long,song2007long,yanai2004new,henderson2008generalized,vydrov2006assessment,rohrdanz2009long,chai2008systematic,chai2008long,lin2013long}
Notably, a systematic search spanning the vast space of possible mathematical forms have been conducted
to find range-separated GGAs with the best general performance.\cite{mardirossian2014exploring} In contrast,
only a few attempts have been made to develop a range-separated meta-GGA functional, i.e., a hybrid model in which
the semilocal part depends not only on the density and density gradient, but also on the kinetic energy density
and in some cases the Laplacian. Empirical functionals of this kind have been proposed by
Lin et al.\cite{lin2012long1,lin2012long2} ($\omega$M05-D and $\omega$M06-D3)
and by Peverati et al.\cite{peverati2011improving} (M11). While these methods are heavily parametrized,
e.g., M11 contains 40 empirical parameters, the available tests show that the improvement over the best
range-separated GGAs is nonuniform and minor.\cite{lin2012long2,mardirossian2015mapping}
A nonempirical range-separated meta-GGA based on the TPSS functional was
tested by Vydrov et al.,\cite{vydrov2006importance} but for thermochemistry this method showed no improvement
over the pure TPSS functional.

The purpose of this work is to construct a reliable range-separated functional in which the short-range
exchange part is a meta-GGA derived from an existing nonempirical semilocal model.

The range-separated exchange energy consists of two components, short-range and long-range,
defined according to the range split of the electron interaction,
\begin{equation}
\frac{1}{s} = \frac{\erfc(\omega s)}{s} + \frac{\erf(\omega s)}{s}, \label{operator-split}
\end{equation}
where $\omega$ is the range separation parameter and $s=|\ra-\rb|$. 
Inserting \eqnref{operator-split} into the definition of the exchange energy yields the formulae for the short-range
and long-range components:
\begin{align}
\exsr=\frac{1}{2}\sum_{\sigma}\iint \frac{\rho_\sigma(\ra) h_\mathrm{X,approx}^\sigma(\ra,\rb) \erfc(\omega s)}{s}\dra \drb, \label{sr-def} \\
\exlr=\frac{1}{2}\sum_{\sigma}\iint \frac{\rho_\sigma(\ra) h_\mathrm{X,exact}^\sigma(\ra,\rb) \erf(\omega s)}{s}\dra \drb. \label{hf-lr-def}
\end{align}
The long-range exchange energy $\exlr$ is based on the exact, orbital-dependent HF exchange hole
\begin{equation}
h_\mathrm{X,exact}^\sigma(\ra,\rb) = -\frac{\left|\sum_i^{N_\sigma}\psi_{i\sigma}^*(\ra) \psi_{i\sigma}(\rb)\right|^2}{\rho_\sigma(\ra)}.
\end{equation}
In the definition of the short-range exchange energy $\exsr$, one has to assume
a specific form of the approximate
exchange hole $h_\mathrm{X,approx}^\sigma$. As in the case of the exchange energy density,
the local definition of the exchange hole is not unique. However, the ambiguity disappears
in the system average of the hole.\cite{ernzerhof1998generalized}

In what follows, we present equations for closed-shell systems with $\rho_\alpha = \rho_\beta = \rho / 2$. There is no loss of 
generality because the exchange functional for arbitrary spin polarizations is simply related to its spin-compensated
counterpart by the formula\cite{oliver1979spin}
\begin{equation}
E_\mathrm{X}\left[ \rhoa, \rhob \right] = \frac 1 2 E_\mathrm{X}\left[2 \rhoa\right] + \frac 1 2 E_\mathrm{X} \left[ 2 \rhob \right].
\end{equation}
For clarity, hereafter we skip the spin index in the exchange hole symbol.

There exists a series of range-separated GGAs which employ various levels of exact constraints
in the model exchange hole inserted into the definition of $E_\mathrm{X}^\mathrm{SR}$.

One of the earliest range-separated functionals are those of Iikura, Tsuenda, Yanai, and Hirao
(ITYH),\cite{iikura2001long} who devised a general technique of converting existing GGAs into range-separated hybrids.
The ITYH scheme was employed in several functionals, including LC-BLYP, LC-BOP, LC-PBEOP,
and CAM-B3LYP.\cite{tawada2004long,song2007long,yanai2004new}

The ITYH exchange hole is based on a simple modification of the LDA exchange hole.\cite{iikura2001long} It has the correct
value at $s=0$,
\begin{equation}
h_\mathrm{X,ITYH}(\ra,s=0) = h_\mathrm{X,exact}(\ra,s=0) = -\frac{\rho(\ra)}{2}, \label{exchange-hole-origin}
\end{equation}
and satisfies the energy integral 
\begin{equation}
\frac 1 2 \int \frac{h_\mathrm{X,ITYH}(\ra, s)}{s} 4 \pi s^2 \diff s = \epsxapprox(\ra), \label{exchange-hole-energy}
\end{equation}
where $\epsxapprox$ is the exchange energy density of a given base functional. 
The ITYH hole fails to fulfill two other exact conditions appropriate to a semilocal functional:
the hole normalization\cite{henderson2008generalized}
\begin{equation}
\int h_\mathrm{X,exact}(\ra, s) 4 \pi s^2 \diff s = -1 \label{exchange-hole-norm}
\end{equation}
and the correct second-order short-range expansion of
the spherically-averaged exchange hole at zero current density,\cite{becke1988correlation,becke1996current,lee1987gaussian}
\begin{align}
h_\mathrm{X,exact}(\ra,s) &= -\frac{\rho}{2} - Q s^2 + \ldots, \label{exchange-hole-expansion} \\
Q &= \frac{1}{12} \nabla^2\rho - \frac{1}{6} \tau + \frac{1}{24} \frac{(\nabla\rho)^2}{\rho}, \label{second-order-coeff}
\end{align}
where $\tau$ is the kinetic energy density
\begin{equation}
\tau = 2 \sum_{i=1}^{N_\text{orb}} |\nabla\psi_{i}|^2.
\end{equation}
It should be stressed that \eqnref{second-order-coeff} cannot be satisfied at the GGA level.

Several GGAs have been developed in which the exchange hole obeys more exact conditions than the ITYH model.
The range-separated PBE functionals of Henderson et al.\cite{henderson2008generalized} and of Vydrov et al.\cite{vydrov2006assessment}
satisfy \eqnref{exchange-hole-origin}, \eqnref{exchange-hole-energy}, \eqnref{exchange-hole-norm},
and only approximately \eqnref{second-order-coeff}. Both methods improve over the ITYH model in atomization
energies and barrier heights.\cite{henderson2008generalized}

Still, there is a possibility for going one rung higher than the existing range-separated GGAs.
This work presents a scheme for construction of meta-GGA range-separated exchange functionals which employ
the kinetic energy density and the Laplacian to exactly include the second-order coefficient of \eqnref{second-order-coeff}.
The method allows one to transform an existing GGA or a meta-GGA model into its range separated variant.
The resulting functional depends on the kinetic energy density and the Laplacian even if the base functional does not.

In the following, we begin by deriving the working equations of the new range-separation scheme. Next,
we search for a preferred combination of the base exchange functional and the accompanying
correlation model. Finally, we test the performance of the selected functional on a test set including
thermochemical energy differences, barrier heights, noncovalent interaction energies, and excitation energies.

\section{Theory}
\subsection{Exchange Hole Model}
Our range-separation scheme requires an exchange hole model which integrates to $\epsxapprox$ and
has enough degrees of freedom to satisfy two further conditions: the exact value of $h_\mathrm{X,approx}$
at $s=0$ and the exact coefficient of $s^2$. These prerequisites are satisfied by the generalized Becke-Roussel (BR)
exchange hole.\cite{becke1989exchange,becke2003real} The spherically-averaged generalized BR hole,
\begin{equation}
  \hxaBR(a,b,\mathcal{N};s) = -\mathcal{N} \frac{a}{16 \pi b s} \left[ (a |b - s| + 1)
    e^{-a|b-s|} -(a|b+s| + 1) e^{-a|b+s|}\right], \label{br-hole}
\end{equation}
includes three parameters, $a$, $b$, and $\mathcal{N}$, which we will define by selecting a subset
of three equations from a wider set of possible conditions.
For any $a>0$ and $b>0$, the normalization integral of $\hxaBR$ is
\begin{equation}
\int \hxaBR(a,b,\mathcal{N}; s) 4 \pi s^2 \diff s = -\mathcal{N}.
\end{equation}
In the original BR model, the parameters $a$ and $b$ are fixed by enforcing
the zeroth- and second-order coefficients of \eqnref{exchange-hole-expansion}, and the normalization is set to $-1$,
i.e., $\mathcal{N}=1$. With these definitions satisfied, the original $\hxaBR$ reduces to the exact exchange hole when applied
to the hydrogen atom.\cite{becke1989exchange}

The original definitions of the BR model have to be modified
so that the electrostatic potential generated by $\hxaBR$ corresponds
to the assumed base exchange energy density:
\begin{equation}
  \frac{1}{2} \int_0^\infty \frac{\hxaBR(a,b,\mathcal{N};s)}{s} 4 \pi s^2 \diff s = \epsxapprox. \label{br-energy-condition}
\end{equation}
The formula for the short-range component of $\epsxapprox$ will be given in Section~\ref{sec-sr-exchange}. Following Becke\cite{becke2003real} and Precechtelova et al.,\cite{precechtelova2015design} we enforce Eq.~\ref{br-energy-condition}
at the cost of relaxing the normalization condition. The set of equations defining the parameters
of $\hxaBR$,
\begin{align}
\frac{x - 2}{x^2} \left(e^x - 1 - \frac{x}{2}\right) &= -\frac{6 Q}{\pi \rho^2} \epsxapprox, \label{main-x-eqn} \\
a &= \sqrt{\pi \rho \frac{\left( 2 - 2  e^x + x \right)}{x  \epsxapprox}},  \label{a-eqn} \\
b &= x / a \label{b-eqn}, \\ 
\mathcal{N} &= 4 \pi \rho e^x / a^3, \label{n-eqn}
\end{align}
is to be solved at each point of space.
(For the derivation of Eqs.~\ref{main-x-eqn}--\ref{n-eqn} see the Appendix of ref~\citenum{becke2003real}.)
For any physically allowed right-hand side, a unique $x > 0$ solves Eq.~\ref{main-x-eqn}. The solution
can be obtained with a numerical solver or interpolation. 

The resulting exchange hole integrates to the given $\epsxapprox$ (\eqnref{br-energy-condition}),
has the exact value at the origin (\eqnref{exchange-hole-origin}), and recovers
the exact coefficient of $s^2$ (\eqnref{second-order-coeff}). However, its normalization integral
differs in general from the exact value of $-1$.

\subsection{Short-Range Exchange Energy} \label{sec-sr-exchange}
The short-range exchange energy density $\epsilon_\mathrm{X,approx}^\mathrm{SR}$ is the difference between
the full-range semilocal exchange and its long-range part:
\begin{equation}
\epsilon_\mathrm{X,approx}^\mathrm{SR} = \epsxapprox - \epsilon_\mathrm{X,approx}^\mathrm{LR}.
\end{equation}
We define $\epsilon_\mathrm{X,approx}^\mathrm{LR}$ using the potential generated by $\hxaBR$:
\begin{equation}
\epsilon_\mathrm{X,approx}^\mathrm{LR} = \frac{1}{2} \int_0^\infty \frac{\hxaBR(s)\erf(\omega s)}{s} 4 \pi s^2 \diff s = \frac 1 2 U_\mathrm{X,approx}^\mathrm{LR}. \label{uxslr-def}
\end{equation}
The integration in \eqnref{uxslr-def} can be done analytically, giving
\begin{align}
  U_\mathrm{X,approx}^\mathrm{LR} &= -\frac{\mathcal{N} \omega}{\bbar}  \erf\left(\bbar\right) \nonumber \\
  &+ \frac{\mathcal{N} \omega}{2
    \bbar} \left(1-\abar^2+\abar\bbar\right)
  \erfc\left(\abar-\bbar\right) \exp\left(\abar^2 - 2 \abar \bbar\right)\nonumber \\
  & + \frac{\mathcal{N} \omega}{2 \bbar} \left(-1+\abar^2+\abar\bbar\right) \label{uxlr}
  \erfc\left(\abar+\bbar\right) \exp\left(\abar^2 + 2 \abar
  \bbar\right), \\
  \abar &= \frac{a}{2 \omega}, \\
  \bbar &= b \omega.
\end{align}
For small values of $\nu$, the right-hand side of \eqnref{uxlr} should be evaluated using a Taylor series expansion
to avoid numerical errors. Finally, the short-range exchange energy is obtained by integrating $\epsilon_\mathrm{X,approx}^\mathrm{SR}$
over the whole space:
\begin{equation}
E_\mathrm{X,approx}^\mathrm{SR} = \int \epsilon_\mathrm{X,approx}^\mathrm{SR} (\ra) \rho(\ra) \diff\ra. \label{ex-sr}
\end{equation}
The complete range-separated exchange energy is the sum of $E_\mathrm{X,approx}^\mathrm{SR}$ and the long-range HF exchange,
\begin{equation}
E_\mathrm{X,approx} = \exsr + \exlr.
\end{equation}

\subsection{One-Electron Self-Interaction Error}
We use the example of the self-interaction error in the ground state of the hydrogen atom 
to illustrate the difference between our meta-GGA range-separation scheme and the existing
GGA approaches.

\begin{figure}
\includegraphics[width=0.70\textwidth]{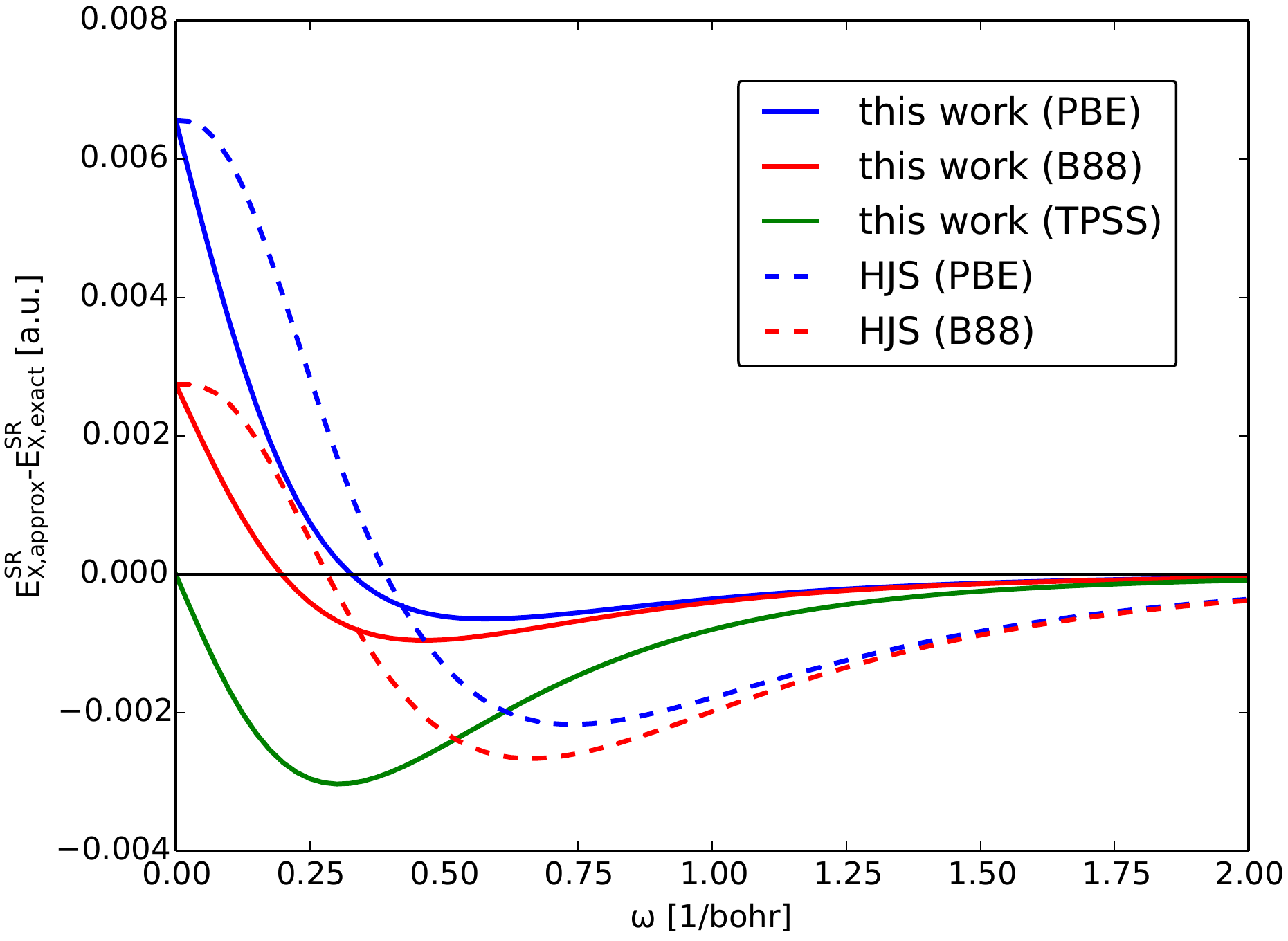}
\captionsetup{justification=RaggedRight}
\caption{Differences between approximate and exact short-range exchange energies of the ground state of the hydrogen atom.
All computations employ the aug-cc-pV5Z basis set\cite{schuchardt2007basis} and HF orbitals.
The short-range GGA models of Henderson et al.\cite{henderson2008generalized,weintraub2009long} are denoted as HJS. 
Correlation energies are not included.} \label{fig-hydrogen-atom}
\end{figure}

The ground state of the hydrogen atom is a difficult limiting case for conventional DFT approximations. 
Using only local variables $\rho(\ra)$ and $\nabla \rho(\ra)$, GGAs have no way of knowing that
the density under consideration belongs to a single-particle system. Therefore, the one-electron
self-interaction error arises as a residual value left by an imperfect cancellation between
an approximate exchange energy and the Coulomb repulsion.\cite{becke1989exchange,perdew1981self} 
A single-electron density can be detected using the kinetic energy density $\tau$, thus meta-GGA 
functionals can, at least partially, reduce the self-interaction error.

The large-$\omega$ behavior of the exact short-range exchange energy of
the hydrogen atom is given by the expansion\cite{gill1996coulomb}
\begin{equation}
E_\mathrm{X,exact}^\mathrm{SR}\left( \omega \rightarrow \infty \right) = -\frac{1}{16 \omega^2} + \frac{1}{32 \omega^4} + \ldots \label{omega-expansion}
\end{equation}
\eqnref{omega-expansion} assumes the exact density. Gill et al.
have shown that the first term on the right-hand side is recovered
already by the local density approximation, but the term of order $1/\omega^4$ requires
$h_\mathrm{X,approx}$ with the correct second-order expansion for small~$s$.\cite{gill1996coulomb}
Indeed, the short-range meta-GGA functionals derived in this work,
which satisfy \eqnref{second-order-coeff}, approach $E_\mathrm{X,exact}^\mathrm{SR}(\omega \rightarrow \infty)$
visibly faster than the existing GGAs (Figure~\ref{fig-hydrogen-atom}). The reduction of errors for large $\omega$
is seen for all tested base functionals: PBE,\cite{perdew1996generalized} B88,\cite{becke1988density}
and TPSS.\cite{tao2003climbing}

\begin{figure}
\includegraphics[width=0.70\textwidth]{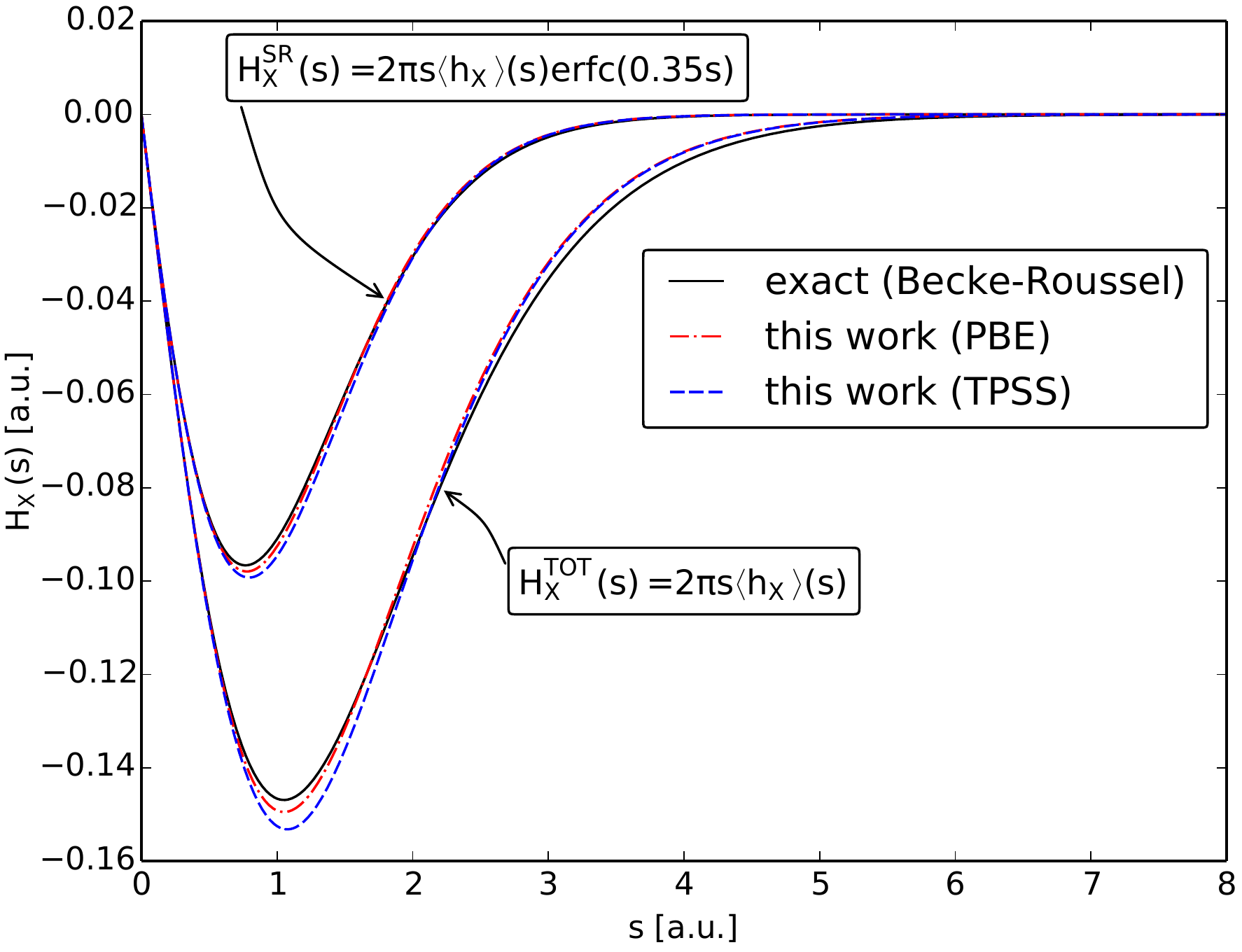}
\captionsetup{justification=RaggedRight}
\caption{Real-space analysis of the contributions to the (short-range) exchange energy of the hydrogen atom.
All computations employ the aug-cc-pV5Z basis set\cite{schuchardt2007basis} and HF orbitals.} \label{fig-hydrogen-xhole}
\end{figure}

Figure~\ref{fig-hydrogen-xhole} shows why, in our scheme, TPSS is not a preferred candidate
for the base exchange functional, and PBE should be used instead.
Let $\langle h_\mathrm{X} \rangle(s)$ denote the system and spherical average
of the exchange hole for the hydrogen atom,
\begin{equation}
\langle h_\mathrm{X} \rangle(s) = \int \rho(\ra) h_\mathrm{X}\left(\ra, s\right) \diff \ra.
\end{equation}
The real-space analysis of the total exchange energy is then expressed as
\begin{equation}
E_\mathrm{X} = \int_0^\infty H_\mathrm{X}^\mathrm{TOT}(s) \diff s, 
\end{equation}
where
\begin{equation}
H_\mathrm{X}^\mathrm{TOT}(s) = 2 \pi s \langle h_\mathrm{X} \rangle(s),
\end{equation}
and the short-range exchange energy is 
\begin{equation}
E_\mathrm{X}^\mathrm{SR}(\omega) = \int_0^\infty H_\mathrm{X}^\mathrm{SR}(s)  \diff s,
\end{equation}
where
\begin{equation}
H_\mathrm{X}^\mathrm{SR}(s) = 2 \pi s \langle h_\mathrm{X} \rangle(s) \mathrm{erfc}(\omega s).
\end{equation}
For the TPSS exchange, $H_\mathrm{X}^\mathrm{TOT}(s)$ is too deep around $s=$~1~bohr and too shallow in the tail, but these
two errors perfectly cancel each other to yield the exact $E_\mathrm{X}$ enforced by the construction of the TPSS exchange. However,
the factor $\mathrm{erfc}(\omega s)$ included in the short-range energy cuts off
the tail of $H_\mathrm{X}^\mathrm{TOT}(x)$,
thus leaving the relatively large short-range error uncompensated in $E_\mathrm{X}^\mathrm{SR}$. By contrast, in the PBE energy,
the short-range and long-range errors in $H_\mathrm{X}^\mathrm{TOT}(s)$ do not cancel perfectly, but the error at short
range is small, and the factor $\mathrm{erfc}(\omega s)$ enhances the error cancellation in $E_\mathrm{X}^{SR}$.

The single-electron density of the hydrogen atom has been previously utilized as a constraint in the design of several
functionals. The TPSS exchange of Tao et al.\cite{tao2003climbing} and the MVS exchange
of Sun et al.\cite{sun2015semilocal} are parametrized to recover the exact exchange energy in this limit.
The hydrogen atom energy is also included in the training set of the empirical M05-2X functional.\cite{zhao2006design}
Here, we use the single-electron limit to estimate the value of $\omega$ which is most appropriate for the range-separated
exchange energy obtained using our scheme. According to Figure~\ref{fig-hydrogen-atom}, our model of the short-range PBE exchange energy
recovers the exact energy at $\omega=0.33$. Later in the text we will show that this value is nearly optimal
for the atomization energies and barrier heights of small molecules.

Apart from its manifestation in approximate exchange energy functionals, the self-interaction error arises 
as a nonvanishing correlation energy of a single-electron system.
In the case of the pure PBE exchange-correlation functional, the total energy of the hydrogen atom
is only \SI{0.0006}{{a.u.}} lower than the exact energy, but at the same time
the correlation contribution amounts to \SI{-0.006}{{a.u.}} (\SI{-3.8}{kcal/mol}).
This error can be eliminated only at the meta-GGA level. The desired improvement
over the PBE correlation is provided by TPSS.\cite{tao2003climbing,perdew2004meta}
The TPSS correlation is built on the PBE formula, but with one-electron self-interaction terms subtracted.\cite{perdew2004meta}
As a result, TPSS yields exactly zero correlation energy for the hydrogen atom, which we regard as a feature
compatible with our exchange model. We will test the advantage of using the TPSS correlation over PBE for general systems
in the following section.

\subsection{Complete Exchange-Correlation Model}
To fully define our exchange-correlation functional, we have to specify the base exchange
functional together with the accompanying model for correlation. We restrict our search to
two exchange-correlation models only: PBE and TPSS. The choice of
these two functionals reflects our preference for methods with a small number of empirical parameters.
Still, it remains possible to pair our range-separation scheme with formulae including multiple adjustable parameters
and to perform a comprehensive empirical optimization.

Let LC-$XY$ denote a range-separated functional where $X$ is the base model for exchange ($\epsxapprox$
in \eqnref{main-x-eqn}), and $Y$ is the accompanying correlation.
Our search comprises three candidate functionals, LC-PBETPSS, LC-PBEPBE, and LC-TPSSTPSS, applied on a set of atomization energies
(AE6\cite{lynch2003small}) and barrier heights (BH6\cite{lynch2003small}). Each functional is employed with a varying
value of $\omega$. The best method is selected for further tests described in the remainder of this paper.
The AE6 and BH6 benchmarks are representative of 109 atomization energies and 44 barrier heights, respectively,
in the Database/3 collection.\cite{lynch2003small} 

LC-TPSSTPSS is the poorest performing functional, which cannot fully benefit from the addition of
the long-range exact exchange. For this functional, a single value of $\omega$ cannot work well
for both AE6 and BH6: the optimal value for the former set is $\omega$=0.0, i.e., the limit of the pure
TPSS functional, whereas for the latter set $\omega$=0.35 minimizes the mean absolute error (MAE).
A similar behavior of the TPSS range-separated hybrid has been observed by Vydrov et al.\cite{vydrov2006importance}
The numerical data for LC-TPSSTPSS are available in the Supporting Information.

The problem of choosing a universally applicable value of $\omega$ arises again in the case of
the candidate based entirely on the PBE model, LC-PBEPBE, albeit it is not as severe
as for LC-TPSSTPSS. At $\omega=0.30$, the average error in the
barrier heights is only 1.6~kcal/mol, but at the same time the error for the atomization energies is as high as
10.5~kcal/mol, which is large compared to the existing range-separated functionals.\cite{henderson2008generalized}

\begin{figure}
\includegraphics[width=0.70\textwidth]{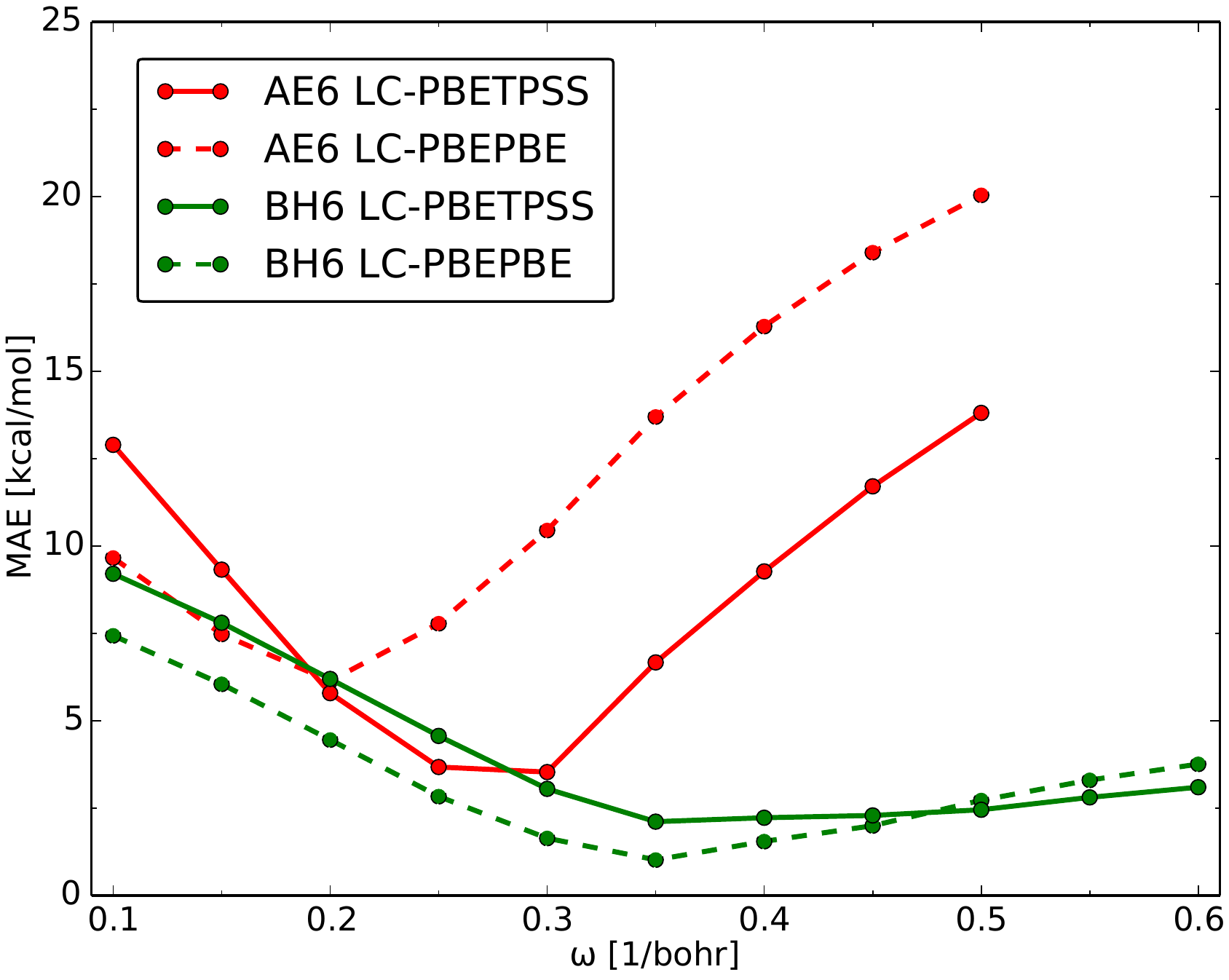}
\captionsetup{justification=RaggedRight}
\caption{Mean absolute errors on the AE6 and BH6 sets.\cite{lynch2003small}
All DFT computations employ the def2-QZVPP basis set.\cite{weigend2005balanced,schuchardt2007basis}
The reference values are taken from ref~\citenum{peverati2014quest} (AE6) and ref~\citenum{karton2008highly} (BH6).} \label{fig-ae6bh6}
\end{figure}

The best overall accuracy is achieved by LC-PBETPSS (Figure~\ref{fig-ae6bh6}).
The optimal range-separation parameter for this functional is in the interval $0.30 \leq \omega \leq 0.35$, depending on
the weight of the BH6 set relative to AE6. (The percentage errors on the BH6 set are much larger
than on AE6, see the Supporting Information.) This result matches our theoretical
estimate, $\omega=0.33$, based on the minimization of the self-interaction error for the hydrogen atom.
Taking into account the relatively large errors in the barrier heights, we choose $\omega=0.35$ for
the final version of LC-PBETPSS recommended for general use. The MAEs at this value of the range-separation
parameter are 6.7~kcal/mol for AE6 and 2.1~kcal/mol for BH6.
LC-PBETPSS is the final, recommended functional which we will employ in the full test set.

The long-range correction proposed here should not be confused with
the correction based on the ITYH scheme, which can be applied, e.g., in the Gaussian program,
to any pure functional. Let us denote by LC-PBETPSS(ITYH) a functional which employs the ITYH-based range-separated PBE
exchange.\cite{iikura2001long} Using the above-described procedure for optimizing the range-separation parameter,
we find that $\omega=0.7$ is optimal simultaneously for AE6 (MAE=$\SI{14.7}{kcal/mol}$) and BH6 (MAE=$\SI{2.6}{kcal/mol}$). For both sets,
LC-PBETPSS(ITYH) is inferior to LC-PBETPSS, but the the difference is especially large for the atomization energies. On the AE6 set,
LC-PBETPSS(ITYH) is only slightly more accurate than the pure PBETPSS functional without any addition of the HF exchange.
For $0.20 \leq \omega \leq 0.35$, where LC-PBETPSS performs well for AE6, LC-PBETPSS(ITYH) yields extremely
large MAEs above 30~kcal/mol. Alternatively, one could combine the range-separated
PBE exchange of Henderson et al.\cite{henderson2008generalized} and the TPSS correlation to obtain LC-PBETPSS(HJS).
While this method performs generally better than LC-PBETPSS(ITYH), for its optimal value of $\omega=0.45$, the errors
for AE6 (MAE=9.9~kcal/mol) and BH6 (MAE=2.4~kcal/mol) are both larger than for LC-PBETPSS.
The numerical data for LC-PBETPSS(ITYH) and LC-PBETPSS(HJS) are available in the Supporting Information.

\subsection{Dispersion Correction}
A dispersion correction compensates for the deficiencies of a semilocal DFT approximation in the modeling of long-range
correlation contributions to noncovalent interaction energies. We test the performance of LC-PBETPSS with the D3 correction
of Grimme et al.\cite{grimme2010consistent}
The general form of the atom-pairwise D3 correction is\cite{grimme2010consistent}
\begin{align}
 E_\mathrm{disp}(\text{D3}) &= -\sum_{A>B} \sum_{n=6,8} s_n \frac{C^{AB}_n}{R^n_{AB}} f_\mathrm{damp}^{(n)}\left(R_{AB}\right), \label{grimme-energy} \\
 f^{(n)}_\mathrm{damp}\left(R_{AB}\right) &= \frac{1}{1+6(R_{AB} / (r_n R^{AB}_0))^{-\alpha_n}}, \label{grimme-damp}
\end{align}
where $f_\mathrm{damp}^{(n)}$ is the damping function. The only functional-dependent parameters are
$r_6$ and $s_8$. The $C_6^{AB}$ dipole-dipole coefficients are 
obtained ab initio, tabulated, and interpolated for the effective coordination numbers in the system of interest.
The minimization of the MAE for LC-PBETPSS-D3 on the S22 set of noncovalent systems\cite{jurecka2006benchmark,podeszwa2010improved} for
LC-PBETPSS yields $r_6=0.88971$. The $1/R^8$ term is not included because it does not decrease the MAE
for the training set ($s_8=0$). We employ the original damping function
$f_\mathrm{damp}^{(n)}(R_{AB})$,\cite{grimme2010consistent} which vanishes for $R_{AB} \rightarrow 0$, instead of
the newer Becke-Johnson damping\cite{grimme2011effect} to avoid double counting of the interaction energy at
short range. Optionally, a 3-body term can be added to model the Axilrod-Teller-Muto contribution to the
dispersion energy:\cite{grimme2010consistent}
\begin{equation}
E_\mathrm{disp}^\text{3-body}(\text{D3}) = -\sum_{A>B>C} C_9^{ABC}\frac{\left( 3 \cos \theta_a \cos \theta_b \cos \theta_c + 1 \right)}{\left(R_{AB} R_{BC} R_{CA}\right)^3} f_\mathrm{damp}^{(9)}\left( \overline{R}_\mathrm{ABC} \right),
\end{equation}
where $\theta_a$, $\theta_b$, and $\theta_c$ are angles between the three interacting atoms,
and $\overline{R}_\mathrm{ABC}$ is the geometric mean of the interatomic distances.
The triple-dipole coefficient $C_9^{ABC}$ is approximated as
\begin{equation}
C_9^{ABC} = -\sqrt{C_6^{AB}C_6^{AC}C_6^{BC}}.
\end{equation}
The nonadditive 3-body term is known to be important for large systems.\cite{grimme2012supramolecular}

\section{Results and Discussion}
\subsection{Electronic-Structure Methods}
The functional developed in this work is denoted as LC-PBETPSS. For the clarity of presentation, let us list its
main characteristics which were discussed in the previous sections. The range-separated exchange
combines the meta-GGA short-range PBE exchange and the 100\% HF exchange at long range. The range-separation parameter
of the exchange is fixed at $\omega=0.35$. The TPSS model is used for the correlation term. The LC-PBETPSS functional
is applied with the D3 dispersion correction (LC-PBETPSS-D3) and for some systems without the dispersion term (LC-PBETPSS).
The LC-PBETPSS functional is implemented in the developer version of the Molpro program.\cite{werner2012molpro}

To make a fair presentation of the performance of the new method,
we have assembled a test set of well-established functionals for comparison.
The LC-$\omega$PBE functional of Vydrov and Scuseria\cite{vydrov2006assessment} is a GGA range-separated functional based on
the PBE exchange and PBE correlation. The numerical comparison between
LC-PBETPSS-D3 and LC-$\omega$PBE-D3 probes the cumulative effect of upgrading the short-range
exchange to meta-GGA and removing the one-electron self-interaction error from the correlation.
The M06-2X empirical meta-GGA functional of Zhao and Truhlar\cite{zhao2008m06} is a workhorse of modern computational
chemistry. Even though this functional reproduces a large part of the dispersion energy in the vicinity of equilibrium
separations, adding the D3 correction slightly improves the results in general. M06-2X-D3 is the best dispersion-corrected meta-GGA hybrid
on the GMTKN30 database.\cite{goerigk2011thorough} $\omega$B97XD is an empirical,
dispersion-corrected, range-separated GGA functional of Chai and Head-Gordon.\cite{chai2008long} It is designed for 
thermochemistry, kinetics, and energies of noncovalent systems. $\omega$B97X\cite{chai2008systematic} is a predecessor
of $\omega$B97XD, which is not optimized for use with a dispersion correction. Still, its design makes it suitable
for spectroscopic
properties.\cite{tsai2013assessment} We employ $\omega$B97X in the part of our tests devoted to excitation energies.
M06-L is an empirical meta-GGA functional which does not contain any HF exchange.\cite{zhao2008m06} It is known for
the reliable description of hydrogen-bonded systems.\cite{chan2014performance} Finally, B3LYP-D3 is an example
of a hybrid functional\cite{stephens1994initio} developed in the 1990s, supplemented with the modern D3 correction.

In addition to DFT methods, for ground-state charge-transfer dimers we use the DLPNO-CCSD(T) method,\cite{riplinger2013natural}
which is a low-scaling approximation within the coupled-cluster wave function formalism including connected triples.
The numerical thresholds for DLPNO-CCSD(T) are set at the ``tight'' level defined in Table 1 of ref~\citenum{liakos2015exploring},
as recommended for noncovalent interactions.\cite{liakos2015exploring} The DLPNO-CCSD(T) computations are performed
with the ORCA 3.0.3 program.\cite{neese2012orca}

\subsection{Hydrogen-Bonded Systems}
\begin{figure}
\includegraphics[width=0.70\textwidth]{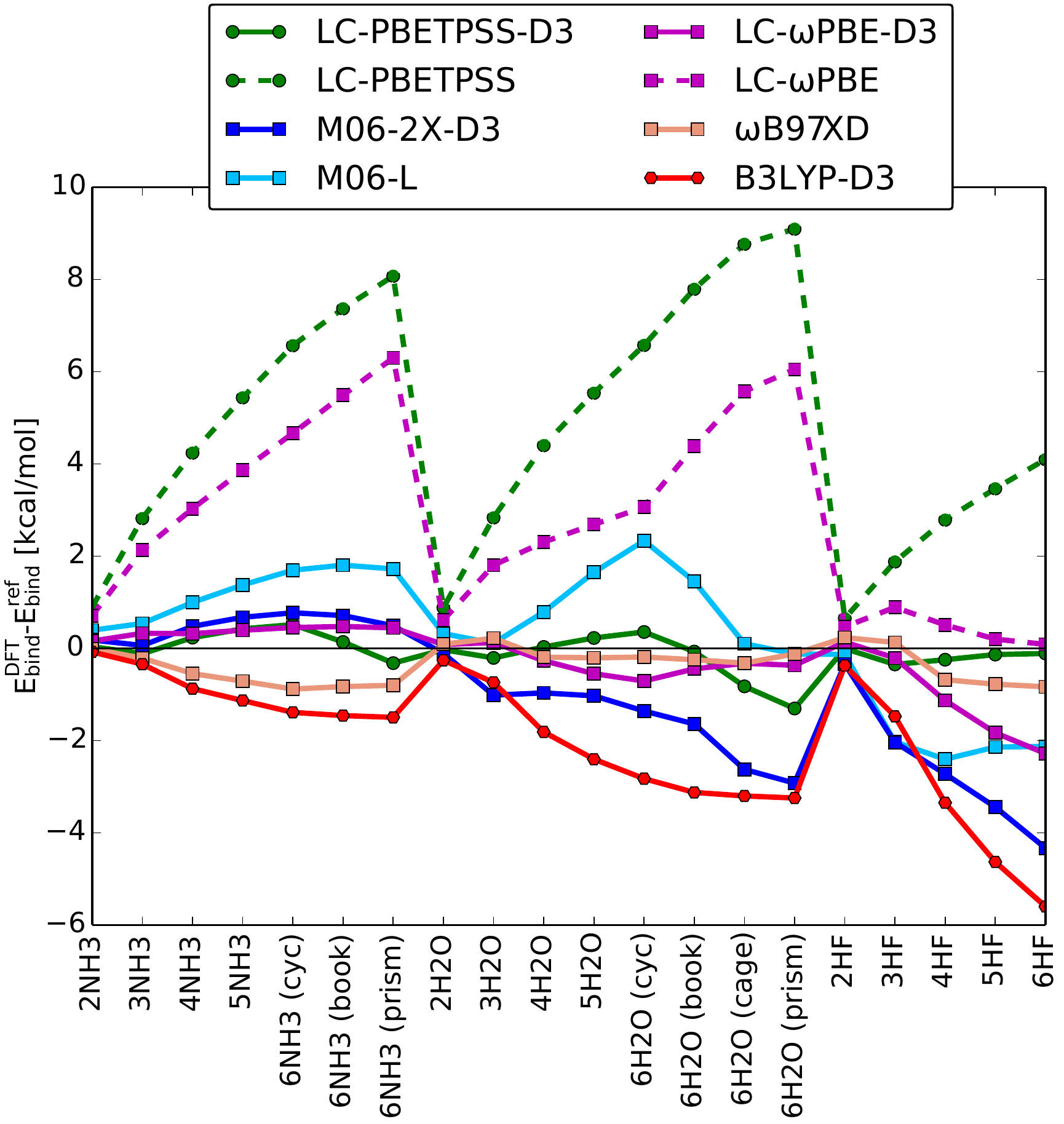}
\captionsetup{justification=RaggedRight}
\caption{Errors for the binding energies of the CEPX33 set. The computational details are provided in Table~\ref{tab-barriers}.}
\label{fig-hb-complexes}
\end{figure}

\begin{figure}
\includegraphics[width=0.70\textwidth]{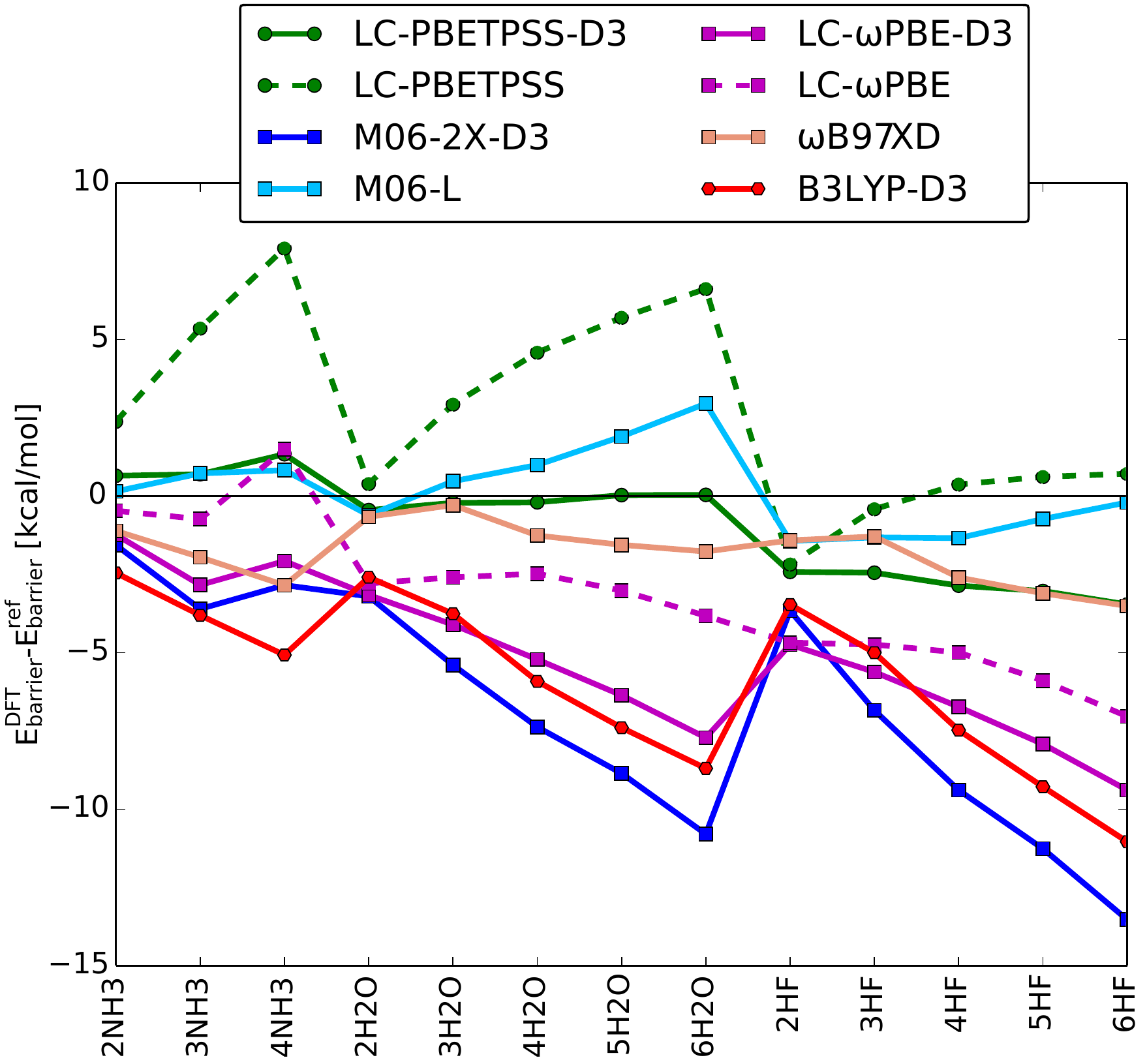}
\captionsetup{justification=RaggedRight}
\caption{Errors for the proton-exchange barriers of the CEPX33 set. The computational details are provided in Table~\ref{tab-barriers}.} \label{fig-hb-barriers}
\end{figure}

Modeling of hydrogen-bonded clusters is still challenging for modern DFT procedures.
Common hybrid GGAs and the M06-type functionals accurately describe the binding energies
but unexpectedly fail for the proton-exchange barriers on the CEPX33 set of \ce{NH3}, \ce{H2O}, and \ce{HF}
clusters.\cite{chan2014performance,karton2012determination} In our tests on the CEPX33 set,
LC-PBETPSS-D3 performs consistently well for both properties
(Figs.~\ref{fig-hb-complexes} and \ref{fig-hb-barriers}). It is the best method for the binding energies
and only slightly less accurate than the best functional (M06-L) for the barriers.
The D3 correction added to LC-PBETPSS improves the results for both binding energies
and barrier heights (Table~\ref{tab-barriers}). This is in contrast to LC-$\omega$PBE, for which the effect
of supplying the dispersion term is inconsistent.

\begin{table}
\centering
  \captionsetup{font=bf,width=0.90\textwidth,justification=RaggedRight}
  \caption{Mean Absolute Errors (kcal/mol) for the Binding Energies (BE) and Proton-Exchange Barriers (PX)
of the CEPX33 Set\tnote{a}}
  \label{tab-barriers}
\begin{threeparttable}
\setlength{\tabcolsep}{7pt}
  \begin{tabularx}{0.4\textwidth}{Xll}
    \toprule
    method &  BE & PX \\
    \midrule
    LC-PBETPSS-D3      &  0.28     & 1.37  \\
    LC-PBETPSS         &  4.71     & 3.09  \\
    M06-L             &  1.21      & 1.05  \\
    $\omega$B97XD     &  0.41     & 1.80 \\
    M06-2X-D3         &  1.40     & 6.79  \\
    LC-$\omega$PBE    & 2.74      & 3.44  \\
    LC-$\omega$PBE-D3 & 0.55      & 5.16  \\
    B3LYP-D3          & 1.99      & 5.84  \\
    \bottomrule
  \end{tabularx}
\begin{tablenotes}
\footnotesize
\item[a] Energies are computed with the aug-cc-pVQZ basis.\cite{schuchardt2007basis} The geometries and reference energies
are taken from ref~\citenum{karton2012determination}.
\end{tablenotes}
\end{threeparttable}
\end{table}

\begin{figure}
\includegraphics[width=0.70\textwidth]{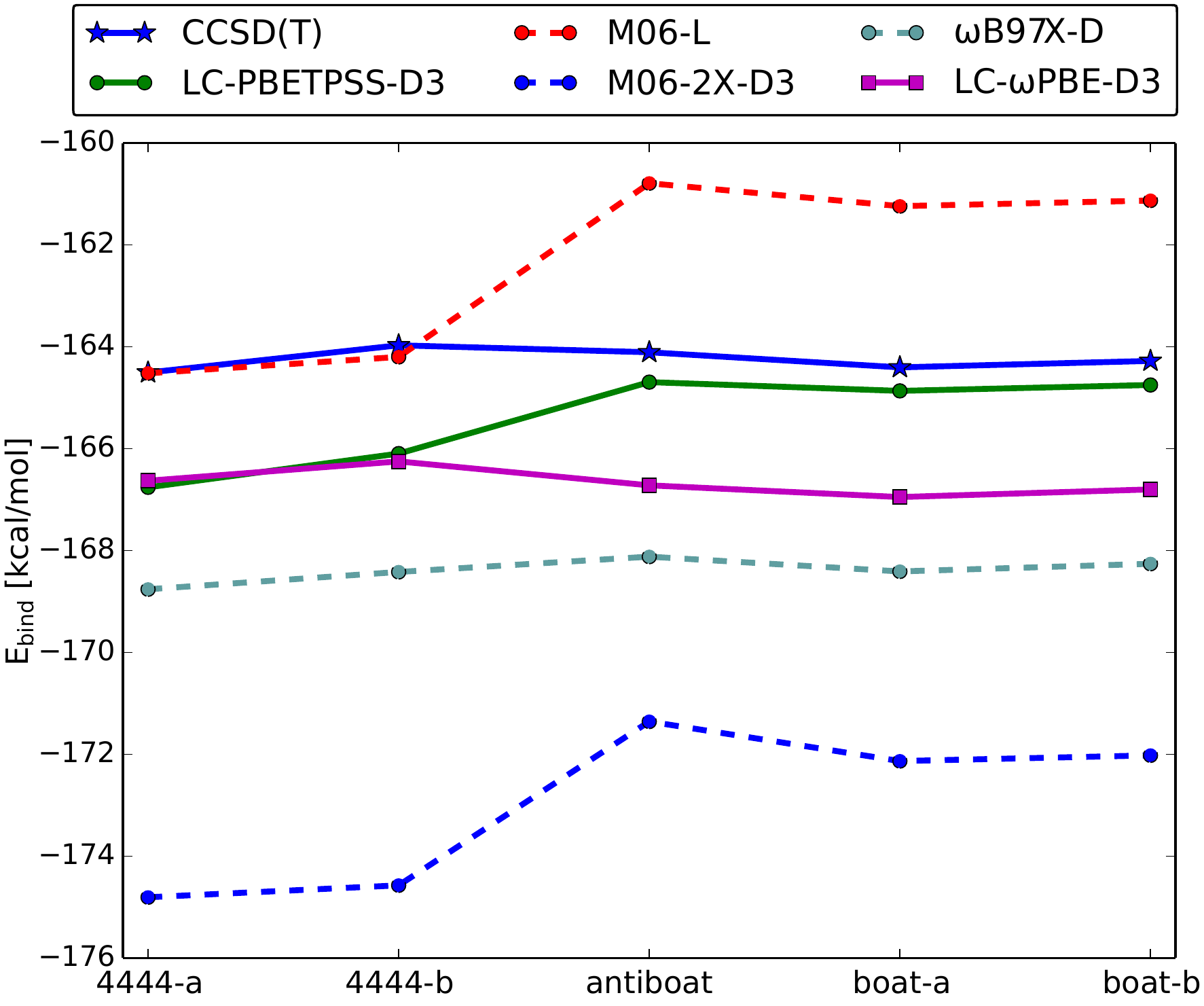}
\captionsetup{justification=RaggedRight}
\caption{Binding energies of water 16-mers. The def2-TZVPPD basis\cite{weigend2005balanced,schuchardt2007basis}
is employed for LC-PBETPSS-D3. The basis-set extrapolated CCSD(T) energies are taken from ref~\citenum{modrzejewski2014range}.
The energies for the existing DFT methods are taken from ref~\citenum{leverentz2013assessing}.} \label{fig-hb-16mers}
\end{figure}

To test if the high accuracy of LC-PBETPSS-D3 persists for systems larger than those of the CEPX33 set,
we apply this functional on the set of water 16-mers studied by Yoo et al.\cite{yoo2010high}
Here, some of the water molecules are connected through hydrogen bonds to four nearest neighbors.
The structures of kind I (4444-a and 4444-b) include eight such nodes,
whereas the structures of kind II (antiboat, boat-a, and boat-b) include four water molecules with such high
connectivity.\cite{yoo2010high}
As illustrated in Figure~\ref{fig-hb-16mers}, LC-PBETPSS-D3 represents reliably the absolute binding energies,
but it predicts that the clusters of kind I are slightly too stable relative to the clusters of kind II.
A similar, yet more pronounced error in the relative energies is present for the M06-type functionals: M06-L
and M06-2X-D3.

\subsection{Noncovalent Charge-Transfer Dimers}
\begin{figure}
\includegraphics[width=0.70\textwidth]{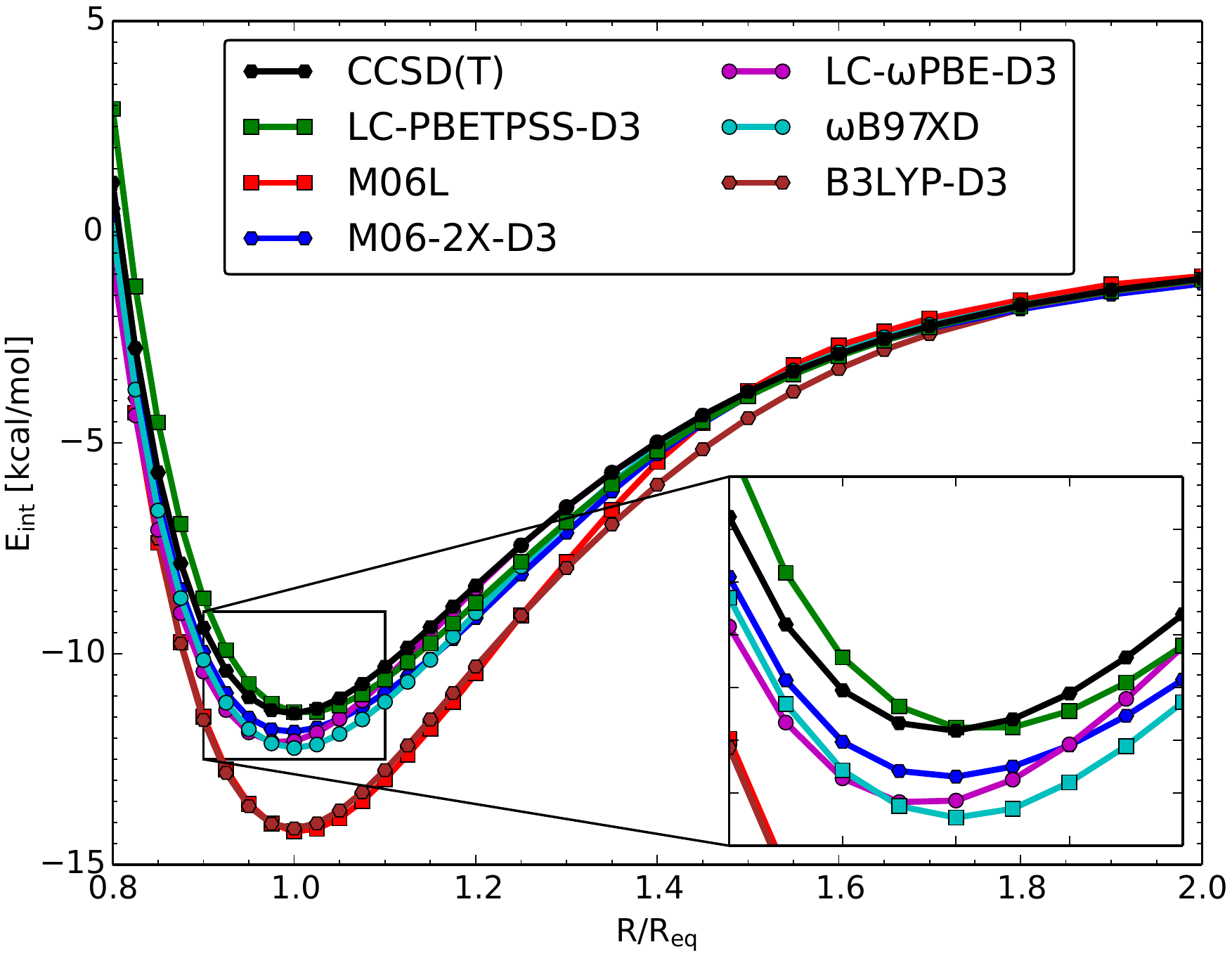}
\captionsetup{justification=RaggedRight}
\caption{Interaction energy curves for the \ce{NH3$...$ClF} dimer.}
\label{fig-nh3clf}
\end{figure}

Since the 1990s, it is known that pure and global hybrid functionals severely overestimate binding energies
of noncovalent charge-transfer dimers.\cite{ruiz1995defining,ruiz1996charge} Range-separated functionals
achieve qualitative improvement by removing the main cause of the overbinding, which is an unrealistic
propensity to transfer electrons between the donor and acceptor. The distinction between range-separated
functionals and more traditional DFT approximations is apparent for the interaction energy curve of
the \ce{NH3$...$ClF} dimer (Figure~\ref{fig-nh3clf}). The two deepest, most overbinding curves
belong to M06-L and B3LYP-D3, a pure functional and a global hybrid, respectively. The range-separated methods,
LC-PBETPSS-D3 in particular, yield a distinct group of energies close to the reference CCSD(T)
curve. The only functional which performs well but is not range-separated, M06-2X-D3, includes
a relatively large fraction of the HF exchange (54\%).

The LC-PBETPSS-D3 curve is extremely close to the reference curve in the vicinity of the equilibrium
separation of \ce{NH3$...$ClF}, but its repulsive part is overestimated. For the compressed dimer
at $R/R_\mathrm{eq}=0.8$, the interaction energy of LC-PBETPSS-D3 ($E_\mathrm{int}=\SI{+2.92}{kcal/mol}$)
is qualitatively different from that of LC-$\omega$PBE-D3 ($E_\mathrm{int}=\SI{-0.72}{kcal/mol}$),
but in accordance with the reference coupled-cluster result ($E_\mathrm{int}=\SI{+1.17}{kcal/mol}$).

Similar behavior of approximate DFT methods is observed for the CT9 set of relatively weakly bound 
donor-acceptor equilibrium dimers (Table~\ref{tab-ct-dimers}). The CT9 set gathers the 
dimers of the CT7/04 set Zhao and Truhlar\cite{zhao2005benchmark} (\ce{C2H2$...$ClF}, \ce{C2H4$...$F2}, \ce{H2O$...$ClF},
\ce{HCN$...$ClF}, \ce{NH3$...$Cl2}, \ce{NH3$...$F2}) and a subset of the complexes studied by 
Yourdkhani et al.\cite{yourdkhani2015interplay} (\ce{CF3CN$...$BF3}, \ce{GeF3CN$...$BF3}, \ce{SiF3CN$...$BF3}).
The MAEs for CT9 are similar for all range-separated functionals and for M06-2X-D3,
but the range-separated hybrids tend to underbind, while M06-2X-D3 predicts
excessive binding. Compared with the uncorrected variants,
both LC-PBETPSS and LC-$\omega$PBE benefit from the D3 dispersion correction.

For additional comparison, we also employ the low-scaling DLPNO-CCSD(T) wavefunction method.
With the MAE of 0.18~kcal/mol on the CT9 set, DLPNO-CCSD(T) is more accurate than any tested DFT method.
However, it is still computationally more expensive than single-determinantal DFT approaches
owing to the relatively strong dependence on the basis set quality.

\begin{table}
\centering
\captionsetup{font=bf,width=0.8\textwidth,justification=RaggedRight}
  \caption{Mean Absolute Errors (kcal/mol) for the Interaction Energies of the CT9 Set of Charge-Transfer Dimers\tnote{a}}
  \label{tab-ct-dimers}
\begin{threeparttable}
  \begin{tabularx}{0.4\textwidth}{Xl}
    \toprule
    method            &  MAE \\
    \midrule
    DLPNO-CCSD(T)     & 0.18      \\
    M06-2X-D3         & 0.37     \\ 
    LC-PBETPSS-D3      & 0.39     \\ 
    LC-PBETPSS        & 1.44     \\
    LC-$\omega$PBE-D3 & 0.41     \\
    LC-$\omega$PBE    & 1.14     \\
    $\omega$B97XD     & 0.41     \\ 
    B3LYP-D3          & 0.73     \\ 
    M06-L             & 0.81     \\  
    \bottomrule
  \end{tabularx}
\begin{tablenotes}
\footnotesize
\item[a] DFT computations are performed with the def2-QZVPP basis. The reference energies at the CCSD(T) level
and the DLPNO-CCSD(T) energies are extrapolated to the basis-set limit (aug-cc-pVTZ $\rightarrow$ aug-cc-pVQZ)
with the automated extrapolation scheme available in ORCA.\cite{neese2012orca} The same computational procedure
is employed for the interaction energy curves of the \ce{NH3$...$ClF} dimer.
\end{tablenotes}
\end{threeparttable}
\end{table}

\subsection{Main-Group Thermochemistry}
To test the performance of LC-PBETPSS-D3 for main-group thermochemistry, we use the sets of
isodesmic reaction energies,\cite{grimme2010alkane} Diels-Alder reaction energies (DARC),\cite{goerigk2011thorough}
and reaction energies with a large contribution of the intramolecular dispersion energy (IDISP).\cite{goerigk2011thorough}

A general-purpose functional has to describe the energy differences between covalently bound structures while
including the contributions from intramolecular noncovalent interactions. A model case of this
kind involves the reaction energies of n-alkane isodesmic fragmentation
\begin{equation}
\ce{CH3(CH2)_{$m$}CH3 + $m$CH4 -> $(m+1)$ C2H6}. \label{isodesmic-reaction}
\end{equation}
Several authors have enumerated the factors which affect the accuracy of approximate DFT for these reactions.
Grimme\cite{grimme2010alkane} noted that a dispersion correction is crucial,
but even a dispersion-corrected semilocal DFT lacks a proper description of middle-range correlation.
Johnson et al.\cite{johnson2012density}
ascribed the size-dependent errors in the reaction energies to the deficient description of regions where the reduced density
gradient changes upon the reaction. An appropriate description of these regions is provided by the PBEsol exchange energy
which obeys the exact second-order expansion for small density gradients.\cite{johnson2012density,csonka2008improved}
Song et al.\cite{song2010calculations} stressed the importance
of correcting the exchange functional via range separation. Finally, Modrzejewski et al.\cite{modrzejewski2014range}
demonstrated a remarkable improvement in the isodesmic reaction energies when using the MCS functional, which
combines the range-separated PBEsol exchange and our meta-GGA correlation optimized to work
with a dispersion correction.\cite{modrzejewski2012first}

\begin{figure}
\includegraphics[width=0.70\textwidth]{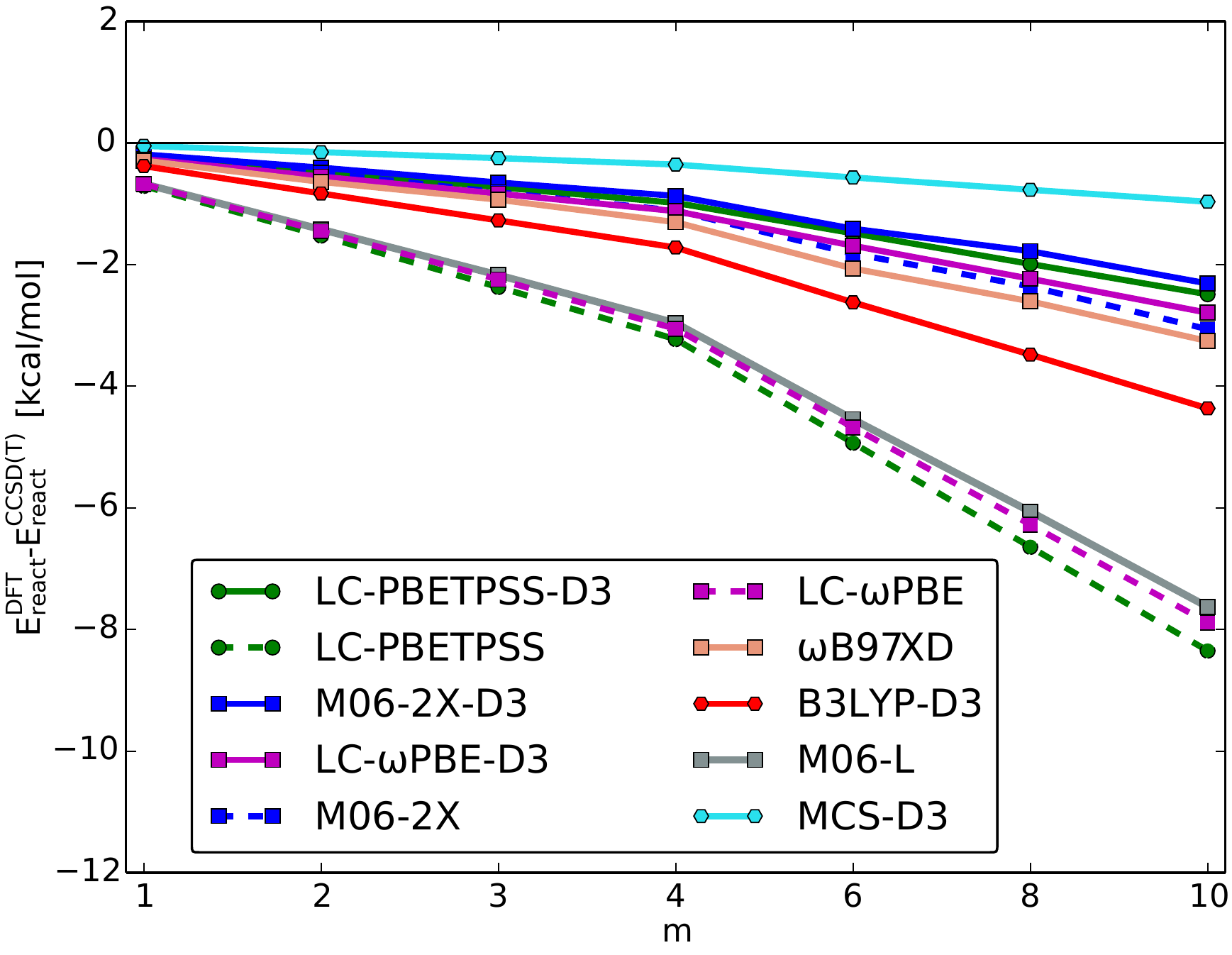}
\captionsetup{justification=RaggedRight}
\caption{Errors in isodesmic reaction energies of n-alkane fragmentation.
The geometries and reference energies at the CCSD(T) level are taken from 
ref~\citenum{grimme2010alkane}. The def2-QZVP basis is employed for all DFT computations
except for MCS-D3.
MCS-D3 is a range-separated functional based on the PBEsol exchange.\cite{modrzejewski2014range}
The energies for MCS-D3 are computed using the def2-TZVPP basis.} \label{fig-isodesmic}
\end{figure}

In our tests, all functionals underestimate alkane stability with the error proportional
to the alkane size (Figure~\ref{fig-isodesmic}). The two error curves with the lowest slope
belong to LC-PBETPSS-D3 and M06-2X-D3. Without the D3 correction, LC-PBETPSS and LC-$\omega$PBE
form a group of outliers together with the pure M06-L functional.
The dispersion term has only a limited effect on M06-2X, which appears to account for the essential
part of the intramolecular dispersion energy via its extensive empirical parametrization.

\begin{table}
\centering
\captionsetup{font=bf,width=0.7\textwidth,justification=RaggedRight}
\caption{Mean Absolute Errors (kcal/mol) for the Reaction Energies of the IDISP and DARC Sets\tnote{a}}\label{tab-idisp-darc}
\begin{threeparttable}
\setlength{\tabcolsep}{4pt}
  \begin{tabularx}{0.50\textwidth}{Xll}
    \toprule
    method                              &  IDISP      & DARC \\
    \midrule
    LC-PBETPSS-D3                       &  2.35\tnote{b}         & 1.38\tnote{c}  \\
    LC-PBETPSS-D3+3body                 &  2.27\tnote{b}         & 1.37\tnote{c}  \\
    LC-PBETPSS                          & 11.38\tnote{b}        &  6.07\tnote{c}  \\
    M06-L\tnote{d}                      &  6.55         & 8.04  \\
    M06-2X-D3\tnote{d}                  &  1.71         & 2.28  \\
    LC-$\omega$PBE-D3\tnote{d}          & 4.13          & 10.04  \\
    LC-$\omega$PBE\tnote{d}             & 8.03          & 6.30  \\
    B3LYP-D3\tnote{d}                   & 6.63          & 10.23  \\
    $\omega$B97XD\tnote{d}              & 2.63          & 1.98  \\
    \bottomrule
  \end{tabularx}
\begin{tablenotes}
\footnotesize
\item[a] Reference energies and geometries are obtained from the companion website of ref~\citenum{goerigk2011thorough}.
\item[b] Computed with the def2-QZVP basis.
\item[c] Computed with the def2-QZVPP basis.
\item[d] Ref~\citenum{goerigk2011thorough}.
\end{tablenotes}
\end{threeparttable}
\end{table}

The DARC subset of the GMTKN30 database\cite{goerigk2011thorough} comprises fourteen Diels-Alder reaction energies
in which the reactants containing multiple conjugated bonds react to form cyclic and bicyclic products
(see Figure~1 in ref~\citenum{johnson2008delocalization}). Most of the existing DFT approximations
underestimate the reaction energies in this set.\cite{johnson2008delocalization}
The reasons for that have general implications for the application of approximate DFT for main group
thermochemistry. Johnson et al.\cite{johnson2008delocalization} have argued that the reactants of the Diels-Alder
reaction have delocalized
electron densities, therefore these structures are artificially stabilized due to the self-interaction (delocalization)
error. On the products side, the bicyclic molecules have bridgehead carbons whose noncovalent repulsion tends
to be overestimated by approximate DFT.\cite{johnson2008delocalization} Because of these two systematic effects,
the energetic gain of going from the reactants to the products is underestimated.

LC-PBETPSS-D3 achieves the lowest mean absolute error of all functionals tested on the DARC set (Table~\ref{tab-idisp-darc}).
The addition of the dispersion correction to LC-PBETPSS reduces the MAE by a factor of four. In contrast, supplying
the D3 term to LC-$\omega$PBE increases the MAE from 6.3~kcal/mol to 10~kcal/mol. The effect of the three-body dispersion term 
included in LC-PBETPSS-D3+3body is negligible due to the small size of the systems.

\begin{figure}
\includegraphics[width=0.70\textwidth]{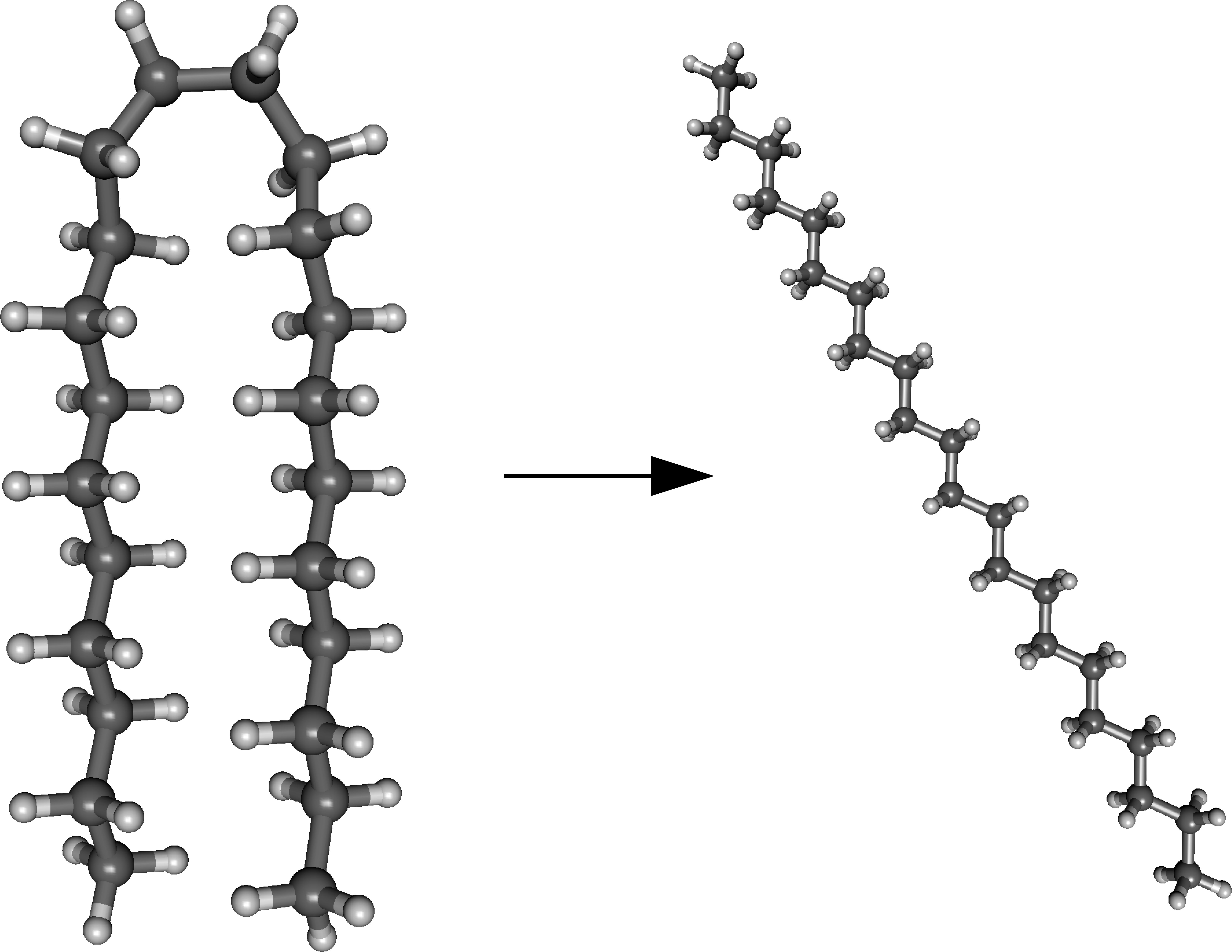}
\captionsetup{justification=RaggedRight}
\caption{Example of a reaction included in the test set for intramolecular dispersion interactions (IDISP).\cite{goerigk2011thorough}}
\label{fig-idisp}
\end{figure}

The IDISP subset of the GMTKN30 database is composed of six reaction energies in which alkanes undergo transformations
between structures with different amounts of the intramolecular dispersion energy.\cite{goerigk2011thorough}
A typical reaction included in IDISP is presented in Figure~\ref{fig-idisp}.
LC-PBETPSS-D3, M06-2X-D3, and $\omega$B97XD are the best methods tested on this set (Table~\ref{tab-idisp-darc}).
The D3 correction is important and beneficial for both LC-PBETPSS and LC-$\omega$PBE. The addition of the three-body D3 term has
a noticeable beneficial effect on the reaction energies predicted by LC-PBETPSS-D3+3body.

\subsection{Excitation Energies}
Numerous authors have reported evidence that there exists a marked advantage of using range-separated functionals
over more traditional DFT approximations for excitation energies of donor-acceptor systems and for Rydberg transitions,
without compromising on valence excitations.\cite{yanai2004new,tawada2004long}
To test the performance of LC-PBETPSS, we apply it to the lowest charge-transfer excitations of
aromatic donor-tetracyanoethylene (Ar-TCNE) pairs
(Table~\ref{tab-ct-excit}) as well as valence and Rydberg excitations
of \ce{CO}, \ce{N2}, \ce{H2CO}, \ce{C2H4}, and \ce{C4H6} (Table~\ref{tab-val-excit}).

Due to the limitations of the software suite in which LC-PBETPSS has been initially implemented,
the excitation energies for this functional are obtained using
real-time time-dependent DFT (RT-TDDFT)
instead of the usual linear response equations.\cite{lopata2011modeling,eshuis2008dynamics}
The propagation of the density matrix was
carried out for \SI{2500}{{a.u.}} (\SI{60}{fs}) for all molecules except for the TCNE-xylene
dimer and ethylene, which were propagated for \SI{3000}{{a.u.}} and \SI{10000}{{a.u.}}, respectively.
The time step in each case was $\Delta t$ = \SI{0.1}{{a.u.}} (\SI{0.0024}{fs}). Each
time a dc pulse with a duration of \SI{0.24}{fs} and field strength of
$E_{\rm max}$ = \SI{0.0001}{{a.u.}} was applied.
All RT-TDDFT calculations were carried out
in the Molpro program.\cite{werner2012molpro}

LC-PBETPSS achieves about the same level of accuracy for Rydberg, valence, and charge-transfer excitations
(Tables~\ref{tab-ct-excit} and \ref{tab-val-excit}). While the
best DFT method for the charge-transfer transitions is $\omega$B97X,\cite{chai2008systematic} there is only
an insignificant difference between $\omega$B97X, LC-PBETPSS, and LC-$\omega$PBE for valence
and Rydberg excitations.

\begin{table}
\captionsetup{font=bf,width=0.70\textwidth,justification=RaggedRight}
\caption{Energies (eV) of the Lowest CT Transitions in Gas-Phase Ar-TCNE Complexes\tnote{a}}
\label{tab-ct-excit}
\begin{threeparttable}
\begin{tabularx}{0.45\textwidth}{X l l l}
\toprule
Ar&benzene&toluene&o-xylene \\
\midrule
ref\cite{hanazaki1972vapor}&3.59&3.36&3.15 \\
$\omega$B97X\cite{chai2008systematic}&3.67&3.34&3.37 \\
LC-$\omega$PBE&4.00&3.65&3.68 \\
LC-PBETPSS&3.87&3.50&3.49 \\
B3LYP&2.06&1.81&1.88 \\
M06-L&1.65&1.46&1.56 \\
M06-2X&3.03&2.93&2.78 \\
GW\cite{blase2011charge}&3.58&3.27&2.89 \\
BNL\tnote{b} & 3.8 & 3.4 & 3.0 \\
\bottomrule
\end{tabularx}
\begin{tablenotes}
\footnotesize
\item[a] DFT calculations employ the cc-pVDZ basis set.\cite{schuchardt2007basis}
\item[b] The range-separated BNL functional\cite{livshits2007well} includes a system-dependent parameter $\omega$. The energies
are taken from ref~\citenum{stein2009reliable}.
\end{tablenotes}
\end{threeparttable}
\end{table}

\begin{table}
\captionsetup{font=bf,width=\textwidth,justification=RaggedRight,skip=5pt}
\caption{Energies (eV) of Valence and Rydberg Transitions in CO, N$_2$, Formaldehyde, Ethylene, and trans-1,3-Butadiene} \label{tab-val-excit}
\begin{threeparttable}
\begin{tabularx}{\textwidth}{X l l l l l l l l l l l}
\toprule
&transition&ref&B3LYP&M06-L&M06-2X&$\omega$B97X\cite{chai2008systematic}&LC-$\omega$PBE&LC-PBETPSS \\
\midrule
\ce{CO}\tnote{a} &$\sigma \rightarrow \pi^*$&8.51\tnote{d}&8.40&8.58&8.22&8.53&8.55&8.66 \\
&$\sigma \rightarrow 3$s&10.78\tnote{d}&9.83&9.35&10.86&10.77&10.84&10.76 \\
&$\sigma \rightarrow 3$p$\sigma$&11.40\tnote{d}&10.21&9.61&10.86&11.22&11.34&11.15 \\
&$\sigma \rightarrow 3$p$\pi$&11.53\tnote{d}&10.27&9.87&10.90&11.31&11.42&11.28 \\
\midrule
\ce{N2}\tnote{a} &$\sigma_g \rightarrow 3$p$\pi_u$&12.90\tnote{d}&11.78&10.85&12.47&12.57&12.68&12.50 \\
&$\sigma_g \rightarrow 3$p$\sigma_u$&12.98\tnote{d}&11.62&10.53&12.53&12.59&12.70&12.52 \\
&$\pi_u \rightarrow 3$s$\sigma_g$&13.24\tnote{f}&12.04&11.76&12.49&12.88&13.01&12.86 \\
\midrule
\ce{H2CO}\tnote{a} & $n \rightarrow 3$sa$_1$&7.09\tnote{d}&6.43&6.14&7.09&7.28&7.26&7.11 \\
&$n \rightarrow 3$pb$_2$&7.97\tnote{d}&7.15&6.49&7.90&8.12&8.11&7.98 \\
&$n \rightarrow 3$pa$_1$&8.12\tnote{d}&7.16&6.57&7.78&8.00&8.00&7.84 \\
&$\sigma \rightarrow \pi^*$&8.68\tnote{d}&9.01&7.01&8.81&8.99&9.11&8.92 \\
\midrule
\ce{C2H4}\tnote{b} &$\pi \rightarrow 3$s&7.11\tnote{e}&6.56&6.60&6.85&7.38&7.52&7.44 \\
&$\pi \rightarrow \pi^*$&7.96\tnote{c}&7.32&7.18&7.47&7.57&7.63&7.69 \\
&$\pi \rightarrow 3$d$\delta$&8.90\tnote{e}&7.61&7.22&8.42&8.98&9.23&9.13 \\
&$\pi \rightarrow 3$d$\delta$&9.08\tnote{e}&7.77&7.47&8.52&9.08&9.33&9.21 \\
&$\pi \rightarrow 3$d$\pi$&9.33\tnote{e}&7.69&7.52&8.58&9.09&9.38&9.28 \\
&$\pi \rightarrow 3$d$\pi$&9.51\tnote{e}&8.09&7.92&8.82&9.46&9.79&9.68 \\
\midrule
\ce{C4H6}\tnote{b} &$\pi \rightarrow \pi^*$&6.32\tnote{c}&5.54&5.62&5.76&5.88&5.97&5.98 \\
&Ryd (2A$_{u}$)&6.66\tnote{e}&5.88&5.87&6.15&6.84&6.94&6.86 \\
&Ryd (2B$_{u}$)&7.07\tnote{e}&6.36&6.09&6.75&7.29&7.40&7.29 \\
&Ryd (3B$_{u}$)&8.00\tnote{e}&6.74&6.39&7.46&8.04&8.30&8.18 \\
\midrule
&MAE&&0.97&1.36&0.42&0.20&0.23&0.22 \\
\bottomrule
\end{tabularx}
\begin{tablenotes}
\footnotesize
\item[a] Energies are computed with the augmented Sadlej basis.\cite{casida1998molecular}
\item[b] Energies are computed with the 6-311(3+,3+)G** basis.\cite{caricato2011oscillator}
\item[c] Theoretical energy at the FCIQMC level, ref~\citenum{daday2012full}.
\item[d] Experimental energy, ref~\citenum{tawada2004long}. 
\item[e] Experimental energy, ref~\citenum{caricato2010electronic}.
\item[f] Experimental energy, ref~\citenum{zhao2006density}.
\end{tablenotes}
\end{threeparttable}
\end{table}

\subsection{Symmetry-Adapted Perturbation Theory}
Symmetry-adapted perturbation theory provides a framework for computation and interpretation of noncovalent
interaction energies.\cite{misquitta2005intermolecular} The energy contributions defined in SAPT
can be computed using approximate functionals, provided that orbital coefficients, orbital energies,
and density response functions are available.

The accuracy of the total interaction energy as well as of the individual SAPT contributions
is contingent on the realistic description of the density tail, therefore
traditional pure and global hybrid functionals must employ asymptotic corrections
of the exchange-correlation potential.\cite{gruning2001shape} Range-separated functionals do not require 
the corrections which change the decay rate of the potential, but they need a procedure
that levels the HOMO energy with negative of the vertical ionization potential (IP).\cite{hapka2014tuned}
The adjustment of the orbital energy involves tuning of the range-separation parameter
for each molecule of interest to satisfy Koopmans' theorem:\cite{stein2009reliable}
\begin{equation}
\epsilon_\mathrm{HOMO}\left(\omega\right) = -\mathrm{IP}(\omega). \label{eqn-omega-optim}
\end{equation}
The procedure of solving Eq.~\ref{eqn-omega-optim} is repeated for each interacting monomer,\cite{hapka2014tuned}
therefore each monomer is assigned its unique value of $\omega$.

To illustrate the importance of using the monomer-dependent range-separation parameters, we employ LC-PBETPSS and
the range-separated PBE functional of Henderson et al.\cite{henderson2008generalized} (HJS-$\omega$PBE)
to compute the total SAPT interaction energies on the A24 set
of noncovalent dimers.\cite{rezac2013describing} Here, the total interaction
 energy is a sum of the first- and second-order SAPT contributions plus a so-called
delta-HF term. Each functional is used to compute the orbital coefficients and energies provided to the SAPT program,
but the exchange-correlation kernel is in every case at the adiabatic local density approximation level.

The improvement of LC-PBETPSS upon using \eqnref{eqn-omega-optim} is clear, with over threefold reduction of 
the MAE for the total interaction energies (Table~\ref{tab-a24-sapt}). The errors are reduced by a similar factor for HJS-$\omega$PBE.
With the monomer-dependent parameter $\omega$, LC-PBETPSS achieves slightly better accuracy
than the common PBE0AC approach, i.e. the PBE0 functional\cite{adamo1999toward}
employed with the asymptotic correction of Gruning et al.\cite{gruning2001shape}

\begin{table}[b]
\centering
\captionsetup{font=bf,width=0.70\textwidth,justification=RaggedRight}
\caption{Mean Absolute Errors (kcal/mol) for the Total SAPT Interaction Energies of the A24 Set\tnote{a}}
  \label{tab-a24-sapt}
\begin{threeparttable}
  \begin{tabularx}{0.40\textwidth}{Xl}
\toprule
method                                    &  MAE \\
\midrule
HJS-$\omega$PBE($\omega$=0.40)            &0.19 \\
HJS-$\omega$PBE($\omega$=$\ast$)\tnote{b} &0.07 \\
LC-PBETPSS($\omega$=0.35)                 &0.30 \\
LC-PBETPSS($\omega$=$\ast$)\tnote{b}      &0.09 \\
PBE0AC&0.12 \\
\bottomrule
\end{tabularx}
\begin{tablenotes}
\footnotesize
\item[a] SAPT calculations employ the aug-cc-pVTZ basis set.
\item[b] Range-separation parameters are adjusted to satisfy \eqnref{eqn-omega-optim}.
\end{tablenotes}
\end{threeparttable}
\end{table}

\section{Summary and Conclusions}
We have proposed a method of creating meta-GGA range-separated exchange functionals
from existing semilocal approximations. Owing to the use of the kinetic energy density
and the Laplacian, the underlying exchange hole has the exact second-order expansion in
the interelectron distance. The importance of this condition is demonstrated for the hydrogenic
density, where the functionals derived using the new approach show a clear reduction of
the self-interaction errors compared to existing range-separated GGAs.

While the method is general, its performance strongly depends on the selected pair of the base
exchange functional and the accompanying correlation. The initial numerical tests on small sets of
atomization energies and barrier heights have shown that the preferred pair of the semilocal
models is the PBE exchange and the TPSS correlation. Therefore, the only functional considered
in the full suite of tests and the method which we recommend for general use is LC-PBETPSS.

The onset of the long-range HF exchange is controlled by the range-separation parameter,
which is estimated theoretically and confirmed by empirical optimization to be $\omega=0.35$.
For applications in SAPT, we recommend to adjust $\omega$ to enforce Koopmans' theorem
for the interacting monomers.

Supplementing LC-PBETPSS with the D3 dispersion correction (comprising only the $1/R^6$ term)
generally improves the accuracy of the method for all test sets considered in this work. We observe additional
slight improvement when a three-body dispersion term is included for large systems.

\begin{figure}
\includegraphics[width=0.70\textwidth]{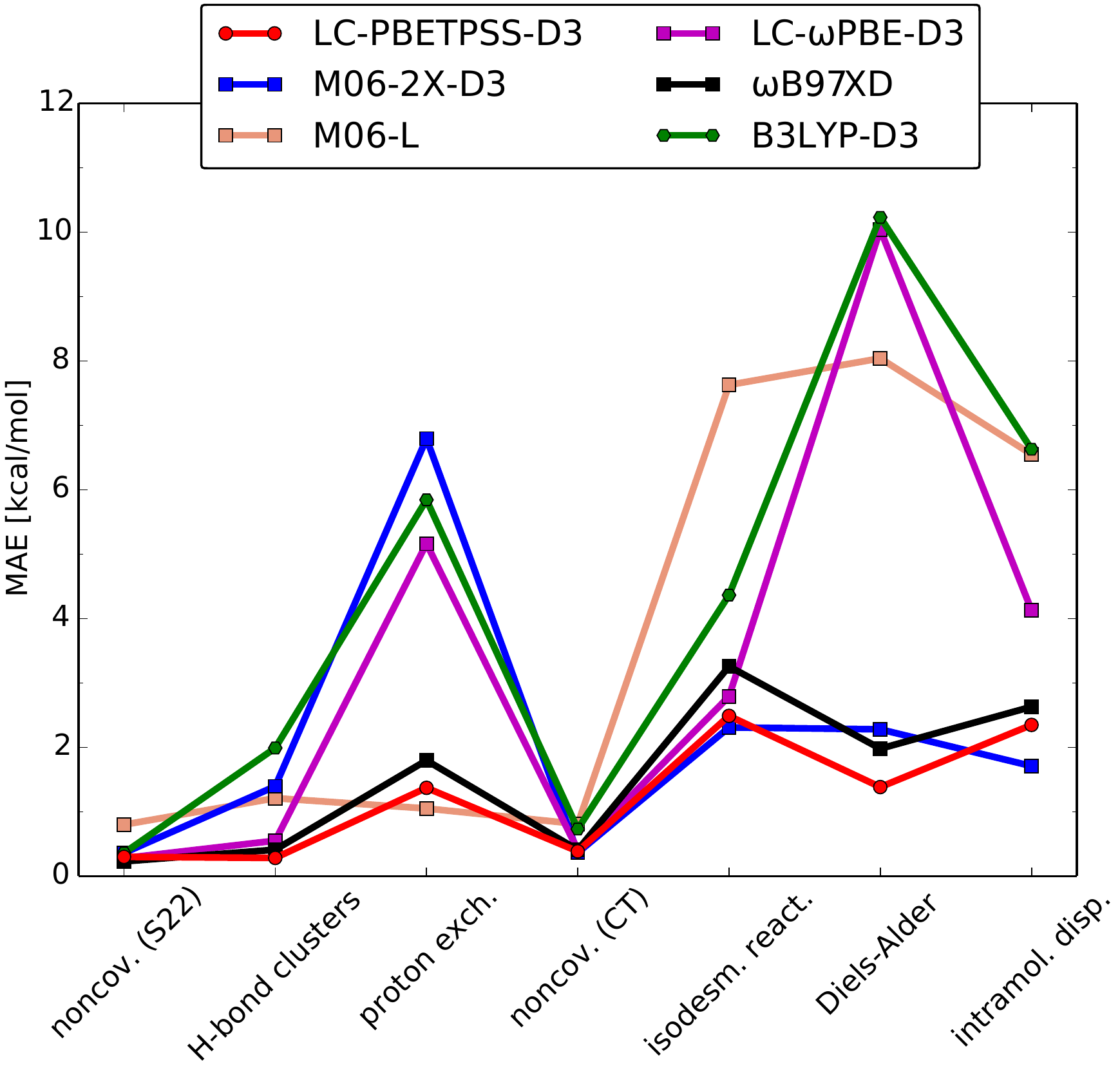}
\captionsetup{justification=RaggedRight}
\caption{General view of the performance of DFT methods on the benchmark sets considered in this work.
The data points for the isodesmic reaction are the absolute errors for dodecane.}
\label{fig-summary}
\end{figure}

As Figure~\ref{fig-summary} illustrates, the accuracy of LC-PBETPSS-D3 is remarkably consistent across
the whole range of tests which probe the performance for noncovalent interaction energies,
barrier heights, and thermochemical energy differences.
The errors corresponding to LC-PBETPSS-D3 are in most cases either the smallest or close
to the best functionals. The only other functional achieving a similar level of consistent
accuracy is $\omega$B97XD. When applied to excited states of small systems, LC-PBETPSS
describes charge-transfer and Rydberg excitations with a similar level of accuracy as
valence excitations.

Compared to LC-$\omega$PBE-D3, the new method offers improved accuracy for the reaction energies of the IDISP set,
Diels-Alder reaction energies, and proton-exchange barriers. While LC-PBETPSS-D3 works better for covalent bonds,
it does not compromise on the accuracy for noncovalent interaction energies. Moreover, the dispersion correction is
more compatible with LC-PBETPSS than with LC-$\omega$PBE. The D3 term is beneficial for LC-$\omega$PBE
when applied to the interaction energies of noncovalent dimers and clusters, but it degrades the accuracy for
proton-exchange barriers and Diels-Alder reaction energies. The performance of LC-PBETPSS-D3 is free
from such irregularities.

Compared to M06-2X-D3, the new method is more reliable for the binding energies and barrier
heights of hydrogen-bonded systems while providing a similar level of accuracy in alkane thermochemistry. 

To conclude, the tests presented in this work show that LC-PBETPSS-D3 combines reliability
with low empiricism. Further work is needed to assess the performance of the new functional
for systems with more complicated electronic structure.

%% \begin{suppinfo}
The Supporting Information is available at DOI: 10.1021/acs.jctc.6b00406.
%Spreadsheets containing detailed numerical data and computational details,
%geometries for the CT9 set and the \ce{NH3$...$ClF} dimer
%% \end{suppinfo}

%\begin{acknowledgement}
\begin{acknowledgments}
M.M. and M.H. were supported by the Polish Ministry of Science and Higher Education
(Grants No.~2014/15/N/ST4/02170 and No.~2014/15/N/ST4/02179). M.M.S. and G.C. were supported
by the National Science Foundation (Grant No.~CHE-1152474).
M.M., M.H., and G.C. gratefully acknowledge additional financial support from the Foundation for Polish Science.
%Special thanks to Aleksandra Tucholska for creating the TOC graphic.
\end{acknowledgments}
%\end{acknowledgement}

\bibliography{biblio}
\end{document}